\documentclass[12pt]{article}
\usepackage{amssymb,amsmath,epsfig}
 \allowdisplaybreaks

\begin{document}
\title{\bf Role of Tilted Congruence and $f(R)$ Gravity on Regular Compact Objects}

\author{Z. Yousaf$^1$ \thanks{zeeshan.math@pu.edu.pk}, Kazuharu Bamba$^2$
\thanks{bamba@sss.fukushima-u.ac.jp} and M. Zaeem-ul-Haq
Bhatti$^1$ \thanks{mzaeem.math@pu.edu.pk}\\
$^1$ Department of Mathematics, University of the Punjab,\\
Quaid-i-Azam Campus, Lahore-54590, Pakistan\\
$^2$ Division of Human Support System,\\ Faculty of Symbiotic
Systems Science,\\ Fukushima University, Fukushima 960-1296, Japan}

\date{}

\maketitle
\begin{abstract}

The purpose of this paper is to check the impact of observer and
Palatini $f(R)$ terms in the formulations of inhomogeneity factors
of spherical relativistic systems. We consider
Lema\^{i}tre-Tolman-Bondi dynamical model as a compact object and
studied its evolution with both tilted and non-tilted observers. We
performed our analysis for particular cases of fluid distribution in
tilted frame and found some energy density irregularity variables.
We found that these variables are drastically different from those
observed by non-tilted observer. The conformal flat dust and perfect
matter contents are homogeneous as long as they impregnate vacuum
core. However, this restriction is relaxed, when the complexity in
the fluid description is increased. The radial fluid velocity due to
tilted congruences and Palatini $f(R)$ curvature terms tend to
produce hindrances in the appearance of energy-density
inhomogeneities in the initially regular spherical stellar
populations.
\end{abstract}

\section{Introduction}

The modern cosmology is inferred as the study of geometry and matter
in the universe which leads to new theoretical ideas about the
theories of gravity analogous with current observations. In $1920s$,
it was Friedmann and Lema\^{i}tre who introduced the concept of
expanding universe which gained significance due to Hubble's
observations in $1930s$. We are still uncertain about the dark side
of the universe, which includes dark energy (DE) and dark matter
(DM), but have confidence that it has some crucial role in
astrophysics and cosmology. The demonstration of the current
accelerated phase of the evolutionary universe proved the dominance
of DE with immense negative pressure \cite{1}. An arguable
alternative to DE is the generalization of general relativity (GR)
which can render cosmic acceleration (for reviews on not only
dark energy problem but also modified gravity theories, see, for example,~\cite{R-DE-MG}).
A simple possible
generalization to GR is the inclusion of non-generic function of
Ricci scalar in the Einstein-Hilber action which can describe the
accelerated expansion and termed as $f(R)$ theory. In deriving the
modified field equations, the procedure which involve the metric and
connections to be independent while performing the variations in the
action is termed as \emph{Palatini approach}. There exists distinct
$f(R)$ models that meet local and cosmological constraints and can
be found in literature \cite{2}. The comprehensive study on the
effectiveness and viability of Palatini approach in $f(R)$ theory of
gravity as compared to observational solar system data has been
presented by Olmo with his coworkers \cite{zb2a}.

Li and Chu \cite{3} presented a framework to study the late-time
cosmic acceleration by constraining the $f(R)$ correction under
Palatini version to GR with high red-shift parameter. Kainulainen
\emph{et al.} \cite{4} analyzed the exterior and interior geometries
of stars by exploring Tolman-Oppenheimer-Volkoff equations in the
background of both Palatini and metric $f(R)$ theory. Fay \emph{et
al.} \cite{5} provided a systematic study for different $f(R)$
gravity models discussing the cosmological dynamics in Palatini
version. Shojai and Shojai \cite{6} studied the features of geodesic
deviation and its congruences by making use of Raychaudhuri's
equation in Palatini $f(R)$ gravity. Sotiriou and Faraoni \cite{7}
surveyed all versions of $f(R)$ gravity from literature and
presented their most significant views comprehensively. Kucukakca
and Camci \cite{8} explored exact solutions with flat
Friedmann-Robertson-Walker (FRW) model for cosmic scale factor by
adopting Noether gauge symmetry approach under Palatini $f(R)$
formalism.

It was enlightened that the universe is not isotropic and
homogeneous at the galactic epochs. To understand the dynamics of
anisotropic and inhomogeneous universe, several cosmological models
have been proposed. Penrose and Hawking \cite{8a} discussed the
irregularities density distribution of spherical relativistic stars
by means of Weyl invariant. Energy density inhomogeneities also
bring forward a crucial role in the process of gravitational
collapse which may lead to appearance of naked singularity
\cite{8aa}. However, the exact relation between the final outcome of
the collapse and density inhomogeneities is still unidentified.
During the evolution of self-gravitating relativistic models, the
role of energy density inhomogeneity have gained much significance
\cite{9}. Herrera \emph{et al.} \cite{9a} discussed the role of
density inhomogeneities on the evolution and structure formation of
spherical anisotropic objects. Recently, the role of Weyl tensor and
super-Poynting vector in various aspects of dissipative and
non-dissipative self-gravitating fluids have been a subject of keen
interest \cite{zb99}.

Mena \emph{et al.} \cite{10} explored the role of inhomogeneity and
anisotropy for a spherically symmetric dust cloud. Di Prisco
\emph{et al.} \cite{10a} looked into non-adiabatic spherically
symmetric collapsing process and explored the role of energy density
inhomogeneities. Chuang \emph{et al.} \cite{11} explored the
possibilities of emergence of inhomogeneities for acceleratory
expanding universe. Herrera \emph{et al.} \cite{11a} studied the
dynamics of dissipative spherical collapse and demonstrated a
relation between density inhomogeneities and Weyl tensor. Herrera
\cite{11aa} formulated inhomogeneity factors for adiabatic and
non-adiabatic relativistic matters and claimed that the system must
satisfy these constraints to achieve stable configurations. Bhatti
and his coworkers \cite{zb11a} examined the impact of extra Ricci
curvature terms on the stability of spherical compact objects filled
with anisotropic relativistic matter distributions. Yousaf \emph{et
al.} \cite{zb11q} explored the contribution of $f(R,T)$ extra
curvature terms in the outcomes of inhomogeneity factors for
dissipative spherical system. They also found factors that causes
the maintenance of homogeneous or inhomogeneous matter state, when
the system departs hydrostatic equilibrium phase in modified gravity
\cite{zb11w}.

A system is said to be tilted if its fluid four-velocity and group
of orbits are not orthogonal and non-tilted otherwise. It is
established in literature that some new interesting results can be
achieved due to tilted observer. The general tilted dynamics of
cosmological models have been considered by Ellis and his
collaborators \cite{12} as well as Bali and his collaborators
\cite{13}. The initial attempt to examine tilted models
qualitatively has been made for Bianchi type II cosmological models
\cite{14}. Pawar \emph{et al.} \cite{15} studied tilted plane
symmetric cosmological models of dissipating isotropic fluid and
investigated that the resulting universe is shearing, expanding and
non-rotating. Apostolopoulos \cite{16} interpreted the dynamical and
geometric features for one class of Bianchi models by presenting the
evolution equations and equilibrium points in tilted and non-tilted
frames. Sahu and Kumar \cite{17} explored the exact solutions for
tilted Bianchi-I cosmological model and examined their geometrical
and physical properties. Sharif and Bhatti \cite{18} explored the
tilted compact objects and developed relationships between tilted
and non-tilted variables which are used in analyzing different
physical quantities. The influences of extra curvature terms coming
from modified gravity on the formulation of inhomogeneity factors
\cite{18inho} and evolution of stellar collapse \cite{18co} have
also been analyzed.

Through the present paper, we explore the inhomogeneity factors
which can control an initially homogeneous system with the evolution
of time. The format of this paper is outlined as follows. In the
next section, we construct all the basic equations by introducing
the concept of tilted observer in the framework of Palatini $f(R)$
theory. In section \textbf{3}, the kinematical quantities, dynamical
as well as evolution equations are explored from the congruence of
tilted observer. Section \textbf{4} is devoted to characterize the
inhomogeneity factors with some particular constraints on the matter
profile. The last section concludes our main findings.

\section{Palatini $f(R)$ Formalism}

The modified gravity theories could be considered as a
powerful tool to understand the enigmatic cosmic evolution.
For Palatini $f(R)$ gravity, the Einstein-Hilbert action is modified as \cite{zb2}
\begin{equation}\label{1}
S_{f(R)}=\frac{1}{2\kappa}\int d^4x\sqrt{-g}f(R)+\mathcal{S}_M,
\end{equation}
where $\kappa$ and $\mathcal{S}_M$ are constant number with
appropriate dimensions, for instance $\kappa=8\pi G$ for GR action and
matter fields action, respectively, while $R\equiv g^{\gamma\delta}R_{\gamma\delta},$
$R_{\gamma\delta}\equiv R^\mu_{~\gamma\mu\delta}$ with $R^\mu_{~\nu\gamma\delta}=\partial_\gamma
\Gamma^\mu_{~\delta\nu}-\partial_\delta
\Gamma^\mu_{~\gamma\nu}+\Gamma^\mu_{~\gamma\sigma}\Gamma^\sigma_{~\delta\nu}
-\Gamma^\mu_{~\delta\sigma}\Gamma^\sigma_{~\gamma\nu}$ indicates Riemann tensor components, the field intensity due to
connections $\Gamma^\mu_{~\gamma\nu}$. As the connection is found dynamically, therefore
one cannot consider $\Gamma_{~\gamma\delta}^\mu=\Gamma_{~\delta\gamma}^\mu$. Due to this reason, we shall keep
$\Gamma_{~\gamma\delta}^\mu\neq\Gamma_{~\delta\gamma}^\mu$ along with $g_{\gamma\delta}=g_{\delta\gamma}$ in our variations.
Varying the above action with $g_{\gamma\delta}$ and
$\Gamma^\mu_{\gamma\delta}$ provide
\begin{align}\label{1n}
\delta S_{f(R)}=\frac{1}{2\kappa}\int d^4x\sqrt{-g}\left[\left(f_R R_{(\gamma\delta)}-\frac{1}{2}g_{\gamma\delta}f\right)
\delta g^{\gamma\delta}+g^{\gamma\delta}f_R\delta R_{\gamma\delta}\right]+\delta \mathcal{S}_M,
\end{align}
where $R_{(\gamma\delta)}$ and $f_R$ indicate symmetric component of the Ricci tensor
and partial Ricci scalar derivation of $f$, respectively. The variations of $R_{\gamma\delta}$ can be expressed as
\begin{align}\label{2n}
\delta R_{\gamma\delta}=\nabla_{\sigma}(\delta\Gamma^\sigma_{~\delta\gamma})-
\nabla_{\delta}(\delta\Gamma^\sigma_{~\sigma\gamma})+2\mathcal{S}^\sigma_{~\mu\delta}\delta\Gamma^\mu_{~\sigma\gamma},
\end{align}
where $\mathcal{S}^\sigma_{~\mu\delta}$ is the torsion tensor defined as $\mathcal{S}^\sigma_{~\mu\delta}=\frac{1}{2}
(\Gamma^\sigma_{~\mu\delta}-\Gamma^\sigma_{~\delta\mu})$. The role of $\delta R_{\gamma\delta}$
quantity in the action (\ref{1}) can be given as
\begin{align}\label{3n}
\int d^4x\sqrt{-g}g^{\gamma\delta}\delta R_{\gamma\delta}&=
\int d^4x[\nabla_{\sigma}(\sqrt{-g}\mathcal{P}^\sigma)+\delta\Gamma^\sigma_{~\delta\gamma}\{2\sqrt{-g}
g^{\gamma\mu}\mathcal{S}^\delta_{~\sigma\mu}\\\nonumber
&+\nabla_\lambda(\sqrt{-g}g^{\gamma\lambda}f_R)
-\nabla_{\sigma}(\sqrt{-g}g^{\gamma\delta}f_R)\}],
\end{align}
where $\mathcal{P}^\sigma=(g^{\gamma\delta}\delta\Gamma^\sigma_{~\gamma\delta}-
g^{\gamma\sigma}\delta\Gamma^\rho_{~\rho\gamma})f_R$. The first term in the above equation
takes the following form
\begin{align}\label{4n}
\nabla_{\sigma}(\sqrt{-g}\mathcal{P}^\sigma)=\partial_\sigma(\sqrt{-g}\mathcal{P}^\sigma)
+\sqrt{-g}f_R[g^{\gamma\delta}S^\lambda_{~\lambda\sigma}-\delta^\delta_{~\sigma}
g^{\gamma\mu}\mathcal{S}^\rho_{~\rho\mu}]\delta\Gamma^\sigma_{\delta\gamma}.
\end{align}
Using these values along with the identity of the surface quantity at the hypersurfaces, i.e., $\int d^4x(\sqrt{-g}\mathcal{P}^\sigma)=0$
in Eq.(\ref{3n}), the field equations can be established as
\begin{align}\label{2}
&f_R(R){R}_{(\gamma\delta)}-[g_{\gamma\delta}f(R)]/2
={\kappa}T_{\gamma\delta},
\end{align}
\begin{align}\label{5n}
&-\nabla_{\sigma}(\sqrt{-g}g^{\gamma\delta}f_R)+\delta^\delta_{~\sigma}\nabla_\lambda(\sqrt{-g}g^{\gamma\lambda}f_R)
+2\sqrt{-g}f_R
(g^{\gamma\delta}\mathcal{S}^\rho_{~\rho\sigma}-\delta^\delta_{~\sigma}g^{\gamma\lambda}\mathcal{S}^\mu_{~\mu\lambda}
+g^{\gamma\mu}\mathcal{S}^\delta_{~\sigma\mu})=\mathcal{H}^{\delta\gamma}_\sigma,
\end{align}
where $\mathcal{H}^{\delta\gamma}_\sigma=-(\delta \mathcal{S}_M/\delta\Gamma^\sigma_{~\delta\gamma})$
and $T_{\gamma\delta}=-(\delta \mathcal{S}_M/\delta g^{\gamma\delta})(2/\sqrt{-g})$. Since we have
considered that the fluid content is not coupled with connection, therefore
$\mathcal{H}^{\delta\gamma}_\sigma=0$. To have a torsionless background, we need to impose
$\mathcal{S}^\delta_{~\sigma\rho}=0.$ In this context, Eq.(\ref{5n}) turns out to be
\begin{align}\label{3}
{\nabla}_\mu(g^{\gamma\delta}\sqrt{-g}f_R({R}))=0,
\end{align}
One can also obtain the similar configurations as mentioned above by removing
the torsionless condition (for details please see \cite{zb1}). On solving Eq.(\ref{3}) (without imposing
torsionless condition, i.e., for the sake of general discussion), we found the relation of connection as follows
\begin{align}\label{6n}
\Gamma^\mu_{\gamma\delta}=\mathcal{C}^\mu_{\gamma\delta}-\frac{2}{3}\mathcal{A}_{\gamma}\delta^\mu_{~\delta},
\end{align}
where
\begin{align}\label{7n}
\mathcal{C}^\mu_{\gamma\delta}=\frac{1}{2}h^{\mu\sigma}(\partial_\gamma h_{\sigma\delta}
+\partial_\delta h_{\sigma\gamma}-\partial_\sigma h_{\gamma\delta}),~~\textrm{with}~~h_{\gamma\delta}=f_Rg_{\gamma\delta}
\end{align}
and $\mathcal{A}_\mu\equiv S^\gamma_{~\gamma\mu}$. Equation (\ref{6n}) has expressed the connection
by means of matter, metric and $\mathcal{A}_\mu$. For torsion-less environment, the quantity
$\mathcal{A}_\mu$ will be zero. Substituting Eq.(\ref{6n}) in Eq.(\ref{3}), it follows that
\begin{eqnarray}\nonumber
&&\frac{1}{f_R}\left({\nabla}_\gamma{\nabla}_\delta-g_{\gamma\delta}
{\Box}\right)f_R+\frac{1}{2}g_{\gamma\delta}{R}+\frac{\kappa}{f_R}
T_{\gamma\delta}+\frac{1}{2}g_{\gamma\delta}\left(\frac{f}{f_R}-R\right)
\\\label{4}
&&+\frac{3}{2f_R^2}\left[\frac{1}{2}g_{\gamma\delta}({\nabla}
f_R)^2-{\nabla}_\gamma f_R{\nabla}_\delta
f_R\right]-\hat{R}_{\gamma\delta}=0,
\end{eqnarray}
where $R=R(g),~R_{\gamma\delta}=R_{\gamma\delta}(g)$ and
${\nabla}_\gamma{\nabla}_\delta f_R$ are calculated through Levi-Civita connection of
the usual metric $g_{\gamma\delta}$. The trace of the above equation can be expressed as
\begin{equation}\label{5}
Rf_R({R})-2f({R})={\kappa}T,
\end{equation}
where $T\equiv g^{\gamma\delta}T_{\gamma\delta}$ is the trace of
usual energy momentum tensor. Equation (\ref{5}) has expressed
Palatini curvature scalar by means of $T$ thereby indicating $R$ and
$f_R$ as the functions of $T$, i.e., $R=R(T)$ and $f_R=f_R(T)$. This
has made their dependence on metric variables, not on independent connections.
The vacuum case, i.e., $T_{\gamma\delta}=0$ would
necessarily leads the differential equation to has a constant
solution that would secure connections to be well-known Levi-Civita.
Further, this would also assign constant value to $f_R$. The
Palatini equation of motion (\ref{4}) can be manipulated as
\begin{equation}\label{6}
{G}_{\gamma\delta}=\frac{\kappa}{f_R}(T_{\gamma\delta}
+{\mathcal{T}_{\gamma\delta}}),
\end{equation}
where
\begin{eqnarray*}\nonumber
{\mathcal{T}_{\gamma\delta}}&=&\frac{1}{\kappa}\left({\nabla}_\gamma{\nabla}_
\delta-g_{\gamma\delta}{\Box}\right)f_R-\frac{f_R}{2\kappa}g_{\gamma\delta}
\left(R-\frac{f}{f_R}\right)\\\nonumber
&+&\frac{3}{2{\kappa}f_R}\left[\frac{1}{2}g_{\gamma\delta}({\nabla}
f_R)^2-{\nabla}_\gamma f_R{\nabla}_\delta f_R\right],
\end{eqnarray*}
while
${G}_{\gamma\delta}~\equiv~{R}_{\gamma\delta}-\frac{1}{2}g_{\gamma\delta}{R}$
is the Einstein tensor,
${\Box}={\nabla}_\gamma{\nabla}_\delta g^{\gamma\delta}$
is a de Alembert operator.

The most general mathematical expression for Lema\^{i}tre-Tolman-Bondi (LTB) spacetime is
\cite{zb3}
\begin{equation}\label{7}
ds^2_-=dt^{2}-\frac{A'^2}{(w+\upsilon)}dr^{2}-C^2(d\theta^2+\sin\theta^2d\phi^2),
\end{equation}
where $\upsilon$ could be $0$ or $\pm1$, $w=w(r)$ following the
constraint $w+\upsilon\geq0$ and prime indicates
$\frac{\partial}{\partial r}$ operator. This spacetime has been used
to study many burning and useful phenomena of our anisotropic and
inhomogeneous universe. It is worthy to mention that on orders much
shorter than Hubble radius, our universe mass density could be
predicted as homogeneous, however, this density regularity is non
longer exists at all scales. One can consider this to be an
applicable scenario for distances larger that $100~Mpc$. The
galactic population has been appeared to be spatially inhomogeneous
for $r$ less than $100Mpc/h$. There has been interesting literature
on this issue \cite{zb4}. Without loss
of generality, one can take $B=A'$ along with $w+\upsilon=1$. Under
this background, the non-static diagonal irrotational LTB metric is
found as follows
\begin{equation}\label{8}
ds_-^2=-dt^2+B^2dr^2+C^2(d\theta^2+\sin\theta^2d\phi^2).
\end{equation}
The geometry of any relativistic celestial bodies is designed by the
gravitational effects coming from its matter source. Such sources
are peculiarly connected with their four-velocities, thus presenting
fluid four-velocities as prominent factors in the formulation of
energy-momentum tensors. The illustrations as well as congruence
kinematics of the gravitational sources could be dissimilar, if the
two feasible relativistic explanations of a given spacetime are
linked through the boost of one of the observer congruences
regarding to the other one. For instance, FRW (with zero curvature)
is a solution of field equation that is coupled with two different
relativistic matter distributions. The first one is ideal fluid and
second is viscous radiating matter source, depending upon the choice
of four-velocity. The former is the solution for those rest
observers who is configuring with reference to time-like congruence,
developed by eigenvectors of $R_{\gamma\delta}$, while the observer
who is moving with relative velocity regarding the first previous
frame will see this to be solution of the later fluid source. Based
upon this concept, we first suppose the comoving coordinate frame,
under which the non-interacting particles have the four-velocity
\begin{equation}\label{9}
u^\gamma=(1,0,0,0),
\end{equation}
with the stress-energy tensor
\begin{equation}\label{10}
T_{\gamma\delta}=\hat{\rho} u_\gamma u_\delta,
\end{equation}
where $\hat{\rho}$ is the energy density. To get tilted congruence,
we assume that fluid distribution has some velocity
$\omega=\omega(r)$ with respect to a new reference frame. Now, we
apply Lorentz boost from locally Minkowskian frame carrying dust
particles to this new frame. Consequently, this gives rise to the
concept of tilted congruences, supported by the following
four-vector field
\begin{equation}\label{11}
U^\gamma=\left(\frac{1}{\sqrt{1-\omega^2}},\frac{\omega}{B\sqrt{1-\omega^2}},0,0\right).
\end{equation}

The fluid corresponding to the tilted frame and vector field
$U_{\gamma}$ is the radiating anisotropic matter distribution, with
the energy-momentum tensor
\begin{equation}\label{12}
T_{\gamma\delta}=({\rho}+{P}_{\bot})U_{\gamma}U_{\delta}+\epsilon
l_\gamma l_\delta-{P}_{\bot}g_{\gamma\delta}+{q}_{\gamma}U_\delta
+({P}_r-{P}_{\bot})S_\gamma S_\delta +{q}_{\delta}U_\gamma,
\end{equation}
where ${\rho},~{q}_\gamma,~\epsilon,~{P}_\bot$ and ${P}_r$ are
energy density, heat flux vector, radiation density, tangential and
radial pressures, respectively. The quantities $S^\gamma$ and
$l^\gamma$ are four-vectors with definitions
\begin{align}\label{13}
S^\gamma&=\left(\frac{\omega}{\sqrt{1-\omega^2}},\frac{1}{B\sqrt{1-\omega^2}},0,0\right),\quad
l^\gamma=\left(\frac{1+\omega}{\sqrt{1-\omega^2}},\frac{1+\omega}{B\sqrt{1-\omega^2}},0,0\right).
\end{align}
The heat flux scalar can be obtained through $S^\gamma$ as
\begin{equation}\label{14}
q^\gamma=qS^\gamma.
\end{equation}
All 4-vectors associated with tilted congruences are satisfying
\begin{equation*}
U^{\gamma}U_{\gamma}=-1=l_\gamma U^\gamma,\quad
S^{\gamma}S_{\gamma}=1=l_\gamma S^\gamma,\quad
l^{\gamma}l_{\gamma}=0=S^\gamma U_\gamma=U^\gamma q_\gamma.
\end{equation*}
For tilted-congruences, we would take $f(R)=R+\frac{\delta^4}{R},$
with $\delta>0$ \cite{zb10}.

\section{Palatini $f(R)$ Ellis Equations}

This section is devoted to develop relationship between Weyl scalar
and LTB dynamical variables widely known as Ellis equations with
\'{a} l\'{a} Palatini formalism. For this purpose, we will formulate
some equations with the help of mass function and field equations.
The Palatini $f(R)$ equations of motion provide the energy variation
of the stellar population gradients respecting time and proximate
surfaces. Through contracted Bianchi identities
\begin{eqnarray}\nonumber
Y^\gamma_{~\delta_;\gamma}=0,~~\textrm{with}~~
Y^\gamma_{~\delta}=T^\gamma_{~\delta}+\mathcal{T}^\gamma_{~\delta},
\end{eqnarray}
and $f(R,T)$ field equations with tilted congruence background can
be found as
\begin{align}\nonumber
&\tilde{\rho}^*+\tilde{\rho}\Theta+\tilde{q}^\dag+\tilde{q}\left\{\omega\Theta
+\frac{\sqrt{1-\omega^2}}{B}\left(\frac{2C'}{C}+\frac{f'_R}{f_R}\right)
+\frac{2\dot{\omega}}{\sqrt{1-\omega^2}}\right\}+\frac{\tilde{\rho}f_R^*}{f_R}\\\label{39}
&+\frac{\tilde{q}f_R^\dag}{f_R}+\frac{\omega
P_\bot'}{B}+P_\bot\left(\Theta+\frac{\dot{f_R}}{f_R}+\frac{\omega
f_R'}{f_R}\right)+\mathcal{D}_0=0,\\\nonumber
&\tilde{P_r}^\dag+a(\tilde{\rho}+\tilde{P_r})+\frac{2\tilde{q}}{3}\left[2\Theta+\sigma-3\omega(\ln
C)^\dag\right]+\tilde{q}^*+\frac{\omega
f_R^*}{f_R}(\tilde{\rho}+P_\bot)-\tilde{q}\sqrt{1-\omega^2}\\\nonumber
&\times
\left(\frac{\dot{B}}{B}+\frac{2\dot{C}}{C}\right)+\frac{1}{f_R\sqrt{1-\omega^2}}\left(\tilde{q}\omega^2\dot{f_R}
-\frac{\tilde{\rho}f'_R}{B}-\frac{P_\bot
f_R'}{B}\right)-\sqrt{1-\omega^2}(P_{\bot}\omega\dot{)}\\\label{40}
&-\frac{\omega^2P_\bot'}{B\sqrt{1-\omega^2}}+\frac{\omega}{\sqrt{1-\omega^2}}
[\dot{\tilde{\mu}}+(\omega\tilde{q}\dot{)}]+\mathcal{D}_1=0,
\end{align}
where $\Theta$ is a Palatini $f(R)$ expansion scalar, $\sigma$ is a
shearing quantity related to shear tensor. Their values for LTB
spacetime are
\begin{align}\label{41}
\Theta&=\frac{1}{\sqrt[3]{1-\omega^2}}\left[\omega\dot{\omega}+\frac{\omega'}{B}+(1-\omega^2)
\left\{\frac{\dot{B}}{B}+\frac{2\dot{f_R}}{f_R}
+\frac{2\dot{C}}{C}+\frac{2\omega C'}{CB} +\frac{\omega
f'_R}{Bf_R}\right\}\right],\\\label{42}
\sigma&=\frac{1}{\sqrt[3]{1-\omega^2}}\left[\omega\dot{\omega}
+\frac{\omega'}{B}+(1-\omega^2)\left\{\frac{\dot{B}}{B}-\frac{\dot{f_R}}{f_R}
-\frac{\dot{C}}{C}-\frac{\omega C'}{CB} +\frac{\omega
f'_R}{Bf_R}\right\}\right],
\end{align}
while $g^\dag=g_{,\mu}S^\mu,~g^*=g_{,\mu}U^\mu$ and
$\mathcal{D}_0,~\mathcal{D}_1$ contain Palatini $f(R)$ dark sector
quantities and are given in Appendix.

The total amount of matter content within the spherical stellar
interior can be found through the well-known Misner-Sharp mass
formalism \cite{zb11}. For metric (\ref{11}), it is given by
\begin{equation}\label{43}
m(t,r)=\frac{C}{2}(1-g^{\gamma\delta}C_{,\gamma}C_{,\delta})
=\left(1+\dot{C}^2 -\frac{C'^2}{B^2}\right)\frac{C}{2}.
\end{equation}
Now, we define an operator which is related to coordinate $r$
derivation as
\begin{eqnarray}\label{44}
D_{C}=\frac{1}{C'}\frac{\partial}{\partial r}.
\end{eqnarray}
Equation (\ref{43}) can be manipulated as
\begin{eqnarray}\label{45}
E\equiv\frac{C'}{B}=\left[1+U^{2}-\frac{2m(t,r)}{C}\right]^{1/2},
\end{eqnarray}
where $U$ denotes fluid velocity which for LTB geometry is found as
$U=\dot{C}$. The $f(R,T)$ field equations and
Eqs.(\ref{43})-(\ref{45}) give
\begin{align}\nonumber
D_Cm&=\frac{\kappa}{2f_R(1-\omega^2)}\left[\tilde{\mu}\left(1+\frac{\omega
U}{E}\right)+\tilde{P}_r\omega\left(\omega+\frac{U}{E}\right)+\tilde{q}\left\{
2\omega+(1+\omega^2)\frac{U}{E}\right\}\right.\\\label{46}
&+\left.(1-\omega^2)
\left(\mathcal{T}_{00}-\frac{U\mathcal{T}_{01}}{EB}\right)\right]C^2,
\end{align}
while the variation of LTB matter content respecting time is
\begin{align}\nonumber
\dot{m}&=\frac{-\kappa}{2f_R(1-\omega^2)}\left[\left\{
(\tilde{\mu}+\tilde{P}_r)\omega+\tilde{q}(1+\omega^2)-\frac{\mathcal{T}_{01}}{B}(1-\omega^2)\right\}E
+\left\{\tilde{\mu}\omega^2\right.\right.\\\label{47}
&\left.\left.+\tilde{P}_r+2\tilde{q}\omega+\frac{\mathcal{T}_{11}}{B^2}(1-\omega^2)\right\}U\right]C^2.
\end{align}
The radial integration of Eq.(\ref{47}) yields
\begin{align}\nonumber
m&=\frac{\kappa}{2}\int_0^C\frac{1}{f_R(1-\omega^2)}\left[\tilde{\mu}\left(1+\frac{\omega
U}{E}\right)+\tilde{P}_r\omega\left(\omega+\frac{U}{E}\right)+\tilde{q}\left\{
2\omega+(1+\omega^2)\right.\right.\\\label{48}
&\times\left.\frac{U}{E}\right\}+\left.(1-\omega^2)
\left(\mathcal{T}_{00}-\frac{U\mathcal{T}_{01}}{EB}\right)\right]C^2dC,
\end{align}
which can be reinterpreted as
\begin{align}\nonumber
\frac{3m}{C^3}&=\frac{3\kappa}{2C^3}\int_0^C\frac{1}{f_R(1-\omega^2)}\left[\tilde{\mu}\left(1+\frac{\omega
U}{E}\right)+\tilde{P}_r\omega\left(\omega+\frac{U}{E}\right)+\tilde{q}\left\{
2\omega+(1+\omega^2)\right.\right.\\\label{49}
&\times\left.\frac{U}{E}\right\}+\left.(1-\omega^2)
\left(\mathcal{T}_{00}-\frac{U\mathcal{T}_{01}}{EB}\right)\right]C^2dC.
\end{align}
After decomposing Weyl tensor into its electric and magnetic parts,
we found that magnetic component turn out to be zero for out LTB
spherical structure. However, its electric part is non-zero. This
can be represented via $U_\gamma$ and $S_\gamma$ as
\begin{equation*}\nonumber
E_{\gamma\delta}=\mathcal{E}\left[S_{\gamma}S_{\delta}-\frac{1}{3}
(g_{\gamma\delta}+U_\gamma U_\delta)\right],
\end{equation*}
with
\begin{align}\label{50}
\mathcal{E}&=\left\{\frac{\ddot{C}}{C}+\left(\frac{\dot{B}}{B}
-\frac{\dot{C}}{C}\right)\frac{\dot{C}}{C}-\frac{\ddot{B}}{B}\right\}
-\left\{\frac{C''}{C}-\left(\frac{C'}{C}+\frac{B'}{B}\right)
\frac{C'}{C}\right\}\frac{1}{2B^{2}}-\frac{1}{2C^{2}}.
\end{align}
This Weyl scalar can be written alternatively via mass function and
field equations as
\begin{align}\label{51}
\mathcal{E}&=\frac{\kappa}{2f_R}\left(\tilde{\mu}-\tilde{P}_r
+P_\bot+\mathcal{T}_{00}-\frac{\mathcal{T}_{11}}{B^2}+\frac{\mathcal{T}_{22}}{C^2}\right)-\frac{3m}{C^3}.
\end{align}
This expression would be very useful to calculate Ellis equation
with tilted congruences in Palatini $f(R)$ gravity.

Now, we are interested to calculate Palatini $f(R)$ distributions of
Ellis equations by following the procedure given by Ellis
\cite{zb12}. These would help us to find irregularity factors in the
energy density of dissipative locally anisotropic matter content
with tilted congruences. Using Eqs.(\ref{46}), (\ref{47}),
(\ref{50}), (\ref{51}) and tilted $f(R,T)$ field equations, these
are formulated as follows
\begin{align}\nonumber
&\left[\mathcal{E}-\frac{\kappa}{2f_R}\left\{\tilde{\mu}-\tilde{P}_r
+P_\bot+\mathcal{T}_{00}-\frac{\mathcal{T}_{11}}{B^2}+\frac{\mathcal{T}_{22}}{C^2}\right\}\right]_{,0}
=\frac{3\dot{C}}{C(1-\omega^2)}\left[-\mathcal{E}+\frac{\kappa}{2f_R}\right.\\\nonumber
&\times\left.\left\{\tilde{\mu}(1+\omega^2)
+P_\bot+\mathcal{T}_{00}+2\tilde{q}\omega
-\frac{\mathcal{T}_{11}}{B^2}\omega^2+\frac{\mathcal{T}_{22}}{C^2}\right\}\right]
+\frac{3\kappa C'}{2BC(1-\omega^2)f_R}\\\label{52}
&\times\left[(\tilde{\mu}+\tilde{P}_r)\omega+\tilde{q}(1+\omega^2)-\frac{\mathcal{T}_{01}}{B}
(1-\omega^2)\right],\\\nonumber
&\left[\mathcal{E}-\frac{\kappa}{2f_R}\left\{\tilde{\mu}-\tilde{P}_r
+P_\bot+\mathcal{T}_{00}-\frac{\mathcal{T}_{11}}{B^2}+\frac{\mathcal{T}_{22}}{C^2}\right\}\right]'
=-\frac{3C'}{C(1-\omega^2)}\left[\mathcal{E}+\frac{\kappa}{2f_R}\right.\\\nonumber
&\times\left.\left\{\tilde{P}_r(1+\omega^2)
-P_\bot-\mathcal{T}_{00}\omega^2+2\tilde{q}\omega
+\frac{\mathcal{T}_{11}}{B^2}-\frac{\mathcal{T}_{22}}{C^2}\right\}\right]
-\frac{3\kappa UC'}{2EC(1-\omega^2)f_R}\\\label{53}
&\times\left[(\tilde{\mu}+\tilde{P}_r)\omega+\tilde{q}(1+\omega^2)-\frac{\mathcal{T}_{01}}{B}
(1-\omega^2)\right].
\end{align}
On considering $f(R)=R$ in above equations, GR Ellis equations for
tilted congruences can be found. However, Ellis equation calculated
by Herrera \emph{et al.} \cite{zb13} can be recovered by taking
$\omega=0$ along with $f(R)=R$.

\section{Inhomogeneities in the Tilted LTB Spheres}

In this section, we would find factors disturbing the energy density
inhomogeneity of the tilted LTB system coupled with anisotropic
dissipative relativistic matter. We would solve modified versions of
Ellis equations that has related the Weyl tensor with fluid source
variables. The study of inhomogeneity parameters occupy enticing
importance in the complete description of stellar gravitational
collapse. The initial homogeneously evolving system will only enter
in the collapsing window once it experiences energy density
irregularities.\\
(i) What factors are actually creating these changes over the
surface of regular relativistic system?\\
(ii) Are dark sector terms affect these inhomogeneity factors?\\
(iii) Furthermore, is this study an observer dependant? \\
In order to answer these issues, we would like to carry out our
analysis with the help of Ellis equations in Palatini $f(R)$
gravity. We shall also check the influence of kinematical parameters
in the modeling of inhomogeneous phases of collapsing stellar
objects. Since, modified gravity may results cumbersome set of
linear equations, therefore, we would like to perform our analysis
by taking simple case of matter source and then we will increase
their order of complexity. We consider following calculations under
the context of current cosmological Ricci scalar constraint. We
shall classify our investigations into couple of streams, i.e.,
radiating/dissipative and non-radiating/non-dissipative populations.

\subsection{Non-Radiating Case}

This section explores inhomogeneity factors of the adiabatic
relativistic tilted matter sources with LTB geometry as
gravitational field in Palatini $f(R)$ gravity. This section
constitutes various non-dissipative choices of matter fields such as
dust, perfect and anisotropic galactic populations, respectively.

\subsubsection{Cloud of Non-interacting Particles}

First we check the geodesic cloud of non-interacting adiabatic
relativistic fragments. So, we consider all pressure gradients,
radiation density as well as heat flux to be zero. For this
subsection, Eqs.(\ref{52}) boils down to
\begin{align}\nonumber
&\left[\mathcal{E}-\frac{\kappa}{2(1-\delta^4R^{-2})}\left\{{\mu}-\frac{\delta^4}{R\kappa}\right\}\right]_{,0}
=\frac{3\dot{C}}{C(1-\omega^2)}\left[-\mathcal{E}+\frac{\kappa}{2(1-\delta^4R^{-2})}
\left\{{\mu}\right.\right.\\\nonumber &\times\left.\left.
(1+\omega^2)-\frac{\delta^4}{R\kappa}\right\}\right]
+\frac{3\kappa{\mu}\omega C'}{2BC(1-\omega^2)(1-\delta^4R^{-2})},
\end{align}
which after using first dynamical equation can be read as
\begin{align}\nonumber
(1-\omega^2)\dot{\mathcal{E}}+\frac{3\dot{C}}{C}\mathcal{E}&=\frac{\kappa}{2(1-\delta^4R^{-2})}
\left[\left\{\frac{3C'}{C}\frac{\omega}{B}-\Theta(1-\omega^2)^{3/2}+\frac{3\dot{C}}{C}\right.\right.\\\label{54}
&\times\left.\left.(1+\omega^2-\delta^4\omega^2)\right\}\mu+(\omega^2-1)\mu'\right].
\end{align}
It is well-known fact that the energy density of dust particles are
regular once the systems impregnate null Weyl scalar. This will
consequently implies zero value of radial derivative of energy
density. Using this result, above equation provides the following
value of expansion scalar
\begin{align}\label{55}
\Theta&=\frac{1}{\sqrt[3]{1-\omega^2}}
\left[\frac{3\dot{C}}{C}(\delta^4\omega^2-\omega^2-1)-\frac{3C'}{C}\frac{\omega}{B}\right].
\end{align}
If irregular system wish to enter in the regular window, its matter
content should need to attain above value of expansion scalar. Now,
the second Ellis equation (\ref{53}) provides
\begin{align}\nonumber
&\left[\mathcal{E}-\frac{\kappa}{2(1-\delta^4R^{-2})}\left\{{\mu}+\frac{\delta^4}{R\kappa}\right\}\right]'
=-\frac{3C'}{C(1-\omega^2)}\left[\mathcal{E}+\frac{\kappa}{2(1-\delta^4R^{-2})}\right.\\\nonumber
&\times\left.\left\{\frac{\delta^4}{R\kappa}\omega^2\right\}\right]
-\frac{3\kappa UC'{\mu}\omega}{2EC(1-\omega^2)(1-\delta^4R^{-2})},
\end{align}
which provides the inhomogeneity condition
\begin{align}\label{56}
&\frac{C'}{C}=\frac{RU\kappa\mu}{3\omega\delta^4E(1-\omega^2)}.
\end{align}
Using above relation in the solution of above Ellis equation with
Schwarzschild radius, i.e., $C=r$, give
\begin{align}\nonumber
\Theta=0.
\end{align}
This shows that homogeneous tilted dust particles with Palatini
$f(R)$ corrections should satisfy expansion-free condition. Under
this condition, the system would experience two very interesting
dynamical process.\\
(i) This condition produces two distinct boundaries (within the
spherical object) in which external one differentiates the
relativistic matter content from the exterior vacuum metric while
the interior one distinguishes central Minkowskian core from the
fluid gravitational source. Under zero expansion scalar, the matter
content evolves without being compressed. For instance, during
expansion of spherical stellar gradient, the changes in its volume
produce similar expansion in the external hypersurface
counterbalancing similar internal surface expansion. Thus, zero
expansion scalar initiates a specific form of system evolution in
such which the inner most shell drags away from the central point
resulting the outcome of vacuum core. Based on this concept,
expansion-free matter populations could be effective for the voids explanation.\\
(ii) The collapsing expansion-free fluid upon approaching towards
the central point experienced shear scalar blowup. The strong
shearing effects cause obstruction in the appearance of apparent
horizon, thereby supporting the existence of naked singularity (NS)
\cite{zb14}. Thus, in nature, NS and expansion-free condition are
weaved together. For the deep understanding of NS appearance,
Virbhadra \emph{et al.} \cite{zb15} developed general formalisms.
Further, Virbhadra and Ellis \cite{zb16} linked this outstanding NS phenomenon
with gravitational lensing and presented some basic foreground
results.

\subsubsection{Locally Isotropic Matter Populations}

Here, we assumed that tilted observer has witnessed that LTB
relativistic metric is designed due to gravitational field produced
by ideal matter sources in Palatini $f(R)$ gravity. Then,
Eq.(\ref{52}) yields
\begin{align}\nonumber
&\left[\mathcal{E}-\frac{\kappa}{2(1-\delta^4R^{-2})}\left\{{\mu}
+\frac{\delta^4}{R\kappa}\right\}\right]_{,0}
=\frac{3\dot{C}}{C(1-\omega^2)}\left[-\mathcal{E}+\frac{\kappa}{2(1-\delta^4R^{-2})}\right.\\\nonumber
&\times\left.\left\{P(1+\omega^2)+
(\mu+P)-\frac{\delta^4}{R\kappa}\omega^2\right\}\right]
+\frac{3\kappa
C'({\mu}+{P})\omega}{2BC(1-\omega^2)(1-\delta^4R^{-2})},
\end{align}
after using Eq.(\ref{39}), above equation provides
\begin{align}\nonumber
&\dot{\mathcal{E}}+\frac{3\mathcal{E}\dot{C}}{C(1-\omega^2)}=\frac{\kappa}{2(1-\delta^4R^{-2})}\left[
\frac{3\dot{C}}{C(1-\omega^2)}-\Theta\sqrt{1-\omega^2}+\frac{3\omega
C'}{BC(1-\omega^2)}\right]\\\label{57}
&\times\left\{\mu-\frac{\alpha(R^2-\delta^4)}{2(1+2\alpha
R)}-\frac{\delta^4}{R\kappa}\right\}-\frac{3\dot{C}\delta^4\omega^2}{2C(1-\omega^2R)(1-\delta^4R^{-2})}
-\frac{\kappa\mu'}{2B(1-\delta^4R^{-2})}.
\end{align}
It is worthy to stress that for comoving system, we have considered
corrections coming from $f(R)=R+{\alpha}R^{2}$ model \cite{zb6,
zb8}, in which $\alpha$ is a positive number. Equation (\ref{57}),
after using some relations between tilted and non-tilted
congruences, provides the following constraint for the existence of
regular energy density with Palatini $f(R)$ background
\begin{align}\label{58}
&\Theta=\frac{3\delta^4\dot{C}\omega^2}{\kappa
CR(1-\omega^2)^{5/2}}\left( \frac{3\dot{C}}{C}+\frac{3\omega
C'}{BC}\right)\left\{\mu-\frac{\alpha(R^2-\delta^4)}{2(1+2\alpha
R)}-\frac{\delta^4}{R\kappa}\right\}^{-1}.
\end{align}
It is evident from the above equation that isotropic LTB stellar
model having Schwarzschild radius will be homogeneous only when the
system impregnates vacuum core. Now, it follows from Eq.(\ref{53})
that
\begin{align}\nonumber
&\left[\mathcal{E}-\frac{\kappa}{2(1-\delta^4R^{-2})}\left\{{\mu}+\frac{\delta^4}{R\kappa}\right\}\right]'
=-\frac{3C'}{C(1-\omega^2)}\left[\mathcal{E}+\frac{\kappa\omega^2}{2(1-\delta^4R^{-2})}\right.\\\nonumber
&\times\left.\left\{{P}-\frac{\delta^4}{R\kappa}\right\}\right]
-\frac{3\kappa
UC'({\mu}+{P})\omega}{2EC(1-\omega^2)(1-\delta^4R^{-2})},
\end{align}
which can be interpreted as
\begin{align}\nonumber
\mathcal{E}+\frac{3C'\mathcal{E}}{C(1-\omega^2)}&=\frac{-3\kappa
C'}{2(1-\omega^2)(1-\delta^4R^{-2})}\left[P\omega^2-\frac{\delta^4\omega^2}{\kappa
R}+\frac{U\omega}{E}(\mu+P)\right]\\\label{59}
&+\frac{\kappa\mu'}{2(1-\delta^4R^{-2})},
\end{align}
from which inhomogeneity factor is found as
\begin{align}\label{60}
\Psi&\equiv-\frac{E}{U}\left(\omega-\frac{U}{E}\right)\left\{
\frac{\alpha(R^2-\delta^4)}{2(1+2\alpha
R)}+\frac{\delta^4}{R\kappa}\right\} -\frac{\delta^2\omega}{R\kappa
E}-\mu.
\end{align}
This shows that when the system is in inhomogeneous phase, it should
need to make null contributions of Weyl scalar and $\Psi$. The major
portion of expression $\Psi$ is controlled by the dark sector terms
coming from Palatini $f(R)$ gravity. Thus, $f(R)$ terms tends to
make hindrance for the same to leave homogeneous as well as
inhomogeneous phases due to their non-attractive nature.

\subsubsection{Locally Anisotropic Gravitational Sources}

For this case, we consider all dissipative terms to be zero in the
first Palatini $f(R)$ Ellis equation. Then, it becomes
\begin{align}\nonumber
&\left[\mathcal{E}-\frac{\kappa}{2(1-\delta^4R^{-2})}\left\{{\mu}-{P}_r
+P_\bot-\frac{\delta^4}{R\kappa}\right\}\right]_{,0}
=\frac{3\dot{C}}{C(1-\omega^2)}\left[-\mathcal{E}+\frac{\kappa}{2(1-\delta^4R^{-2})}\right.\\\nonumber
&\times\left.\left\{{\mu}(1+\omega^2)
+P_\bot-\frac{\delta^4\omega^2}{R\kappa}\right\}\right]
+\frac{3\kappa ({\mu}+{P}_r)\omega
C'}{2BC(1-\omega^2)(1-\delta^4R^{-2})}.
\end{align}
Equation (\ref{39}), after performing some mathematical exercise,
give
\begin{align}\nonumber
&\left\{\mathcal{E}+\frac{\kappa~\Pi}{2(1-\delta^4R^{-2})}\right\}_{,0}
+\frac{3\dot{C}}{C(1-\omega^2)}\left\{\mathcal{E}+\frac{\kappa~\Pi}{2(1-\delta^4R^{-2})}\right\}
=\frac{\kappa\mu}{2(1-\delta^4R^{-2})}\\\nonumber
&\left[\frac{1}{(1-\omega^2)}\left\{\frac{3\dot{C}}{C}(2+\omega^2)-\frac{6\omega
C'}{BC}\right\}-\Theta\sqrt{1-\omega^2}\right]-\frac{3\kappa}{2(1-\delta^4R^{-2})(1-\omega^2)}\\\label{61}
&\left(\frac{\dot{C}}{C}-\frac{\omega
C'}{BC}\right)\frac{\hat{\mu}(R^2-\delta^4)}{R^2(1+2\alpha
R)}+\frac{D_2}{2(1-\delta^4R^{-2})},
\end{align}
with
\begin{align}\nonumber
D_2&=\left\{ \frac{\alpha(R^2-\delta^4)}{2(1+2\alpha
R)}+\frac{\delta^4}{R}\right\}\left[\Theta\sqrt{1-\omega^2}-\frac{6}{(1-\omega^2)}\left(\frac{\dot{C}}{C}-\frac{\omega
C'}{BC}\right)\right].
\end{align}
It can be seen from the literature than the scalar quantity that
control the inhomogeneity emergence in anisotropic sources is the
trace free part of the tensor that came from the orthogonal
splitting of Riemann curvature tensor. Such scalar has been dubbed
as $X_{TF}$. We found that the configurations of  squiggly brackets
terms in first and second mathematical expressions of the above
equation resemble with $X_{TF}$. Using this result, we find that
following value of expansion which the anisotropic stellar
populations must attain to achieve homogeneity in their energy
densities.
\begin{align}\label{62}
\Theta&=\frac{3}{\sqrt[3]{1-\omega^2}}\left[(2+\omega^2)\frac{\dot{C}}{C}-\frac{2\omega
C'}{BC}-\frac{\kappa\hat{\mu}(R^2-\delta^4)}{R^2(1+2\alpha
R)}\left(\frac{\dot{C}}{C}-\frac{\omega
C'}{BC}\right)+\frac{D_2}{3}(1-\omega^2)\right].
\end{align}
The second Ellis equation for anisotropic sources boils down to
\begin{align}\nonumber
&\left[\mathcal{E}-\frac{\kappa}{2(1-\delta^4R^{-2})}\left\{{\mu}-{P}_r
+P_\bot-\frac{\delta^4}{R\kappa}\right\}\right]'
=-\frac{3C'}{C(1-\omega^2)}\left[\mathcal{E}+\frac{\kappa}{2(1-\delta^4R^{-2})}\right.\\\nonumber
&\times\left.\left\{{P}_r(1+\omega^2)
-P_\bot-\frac{\delta^4\omega^2}{R\kappa}\right\}\right]
-\frac{3\kappa ({\mu}+{P}_r)\omega
UC'}{2EC(1-\omega^2)(1-\delta^4R^{-2})}.
\end{align}
This equation after some lengthy but easy mathematical manipulations
yields
\begin{align}\nonumber
&\left\{\mathcal{E}+\frac{\kappa~\Pi}{2(1-\delta^4R^{-2})}\right\}'
+\frac{3C'}{C(1-\omega^2)}\left\{\mathcal{E}+\frac{\kappa~\Pi}{2(1-\delta^4R^{-2})}\right\}
=\frac{\kappa(\mu+D_1)}{2(1-\delta^4R^{-2})}\\\label{63}
&+\frac{\kappa U\omega}{2E(1-\omega^2)(1-\delta^4R^{-2})}
\left\{\mu+\frac{\hat{\mu}(R^2-\delta^4)}{R^2(1+2\alpha R)}\right\},
\end{align}
where
\begin{align}\nonumber
D_1&=\frac{\kappa}{2(1-\delta^4R^{-2})(1-\omega^2)}\left[\frac{3C'\omega^2\delta^4}{CR}+\frac{\omega}{E}
\left\{\frac{2\delta^4}{R\kappa}+\frac{\alpha(R^2-\delta^4)}{R^2(1+2\alpha
R)}\right\}\right].
\end{align}
Here, we also noted the same mathematical combinations as we
observed in Eq.(\ref{61}). Therefore, the regular energy density can
be achieved by the system if the system makes null value to the
following parameter, $\Phi$
\begin{align}\label{64}
\Phi&\equiv\frac{R^2(1-2\alpha
R)}{R^2-\delta^4}\left\{\frac{D_1}{U\omega}(1-\omega^2)+\mu\right\}-\hat{\mu}.
\end{align}
For Schwarzschild radius, Eq.(\ref{62}) provides non-zero value to
expansion scalar. In comparison with previous cases, $\Theta=0$ was
the necessary condition for dust and isotropic fluid systems to
attain homogeneous energy density.

\subsection{Radiating Case}

In this subsection, we explore irregularity factor for the tilted
observer who observed that LTB geometry of the stellar object is
formed due to dissipative dust source. This cloud is dissipating in
the mode of both diffusion and free-streaming approximations.
Therefore, we take all anisotropic pressure gradients to be zero,
then Eqs.(\ref{52}) and (\ref{53}) give
\begin{align}\nonumber
&\left[\mathcal{E}-\frac{\kappa}{2(1-\delta^4R^{-2})}\left\{\tilde{\mu}-\frac{\delta^4}{R\kappa}\right\}\right]_{,0}
=\frac{3\dot{C}}{C(1-\omega^2)}\left[-\mathcal{E}+\frac{\kappa}{2(1-\delta^4R^{-2})}\right.\\\label{65}
&\times\left.\left\{\tilde{\mu}(1+\omega^2) +2\tilde{q}\omega
-\frac{\delta^4\omega^2}{R\kappa}\right\}\right] +\frac{3\kappa
[\tilde{\mu}+\tilde{q}(1+\omega^2)]C'}{2BC(1-\omega^2)(1-\delta^4R^{-2})},\\\nonumber
&\left[\mathcal{E}-\frac{\kappa}{2(1-\delta^4R^{-2})}\left\{\tilde{\mu}-\frac{\delta^4}{R\kappa}\right\}\right]'
=-\frac{3C'}{C(1-\omega^2)}\left[\mathcal{E}+\frac{\kappa}{2(1-\delta^4R^{-2})}\right.\\\label{66}
&\times\left.\left\{2\tilde{q}\omega-\frac{\delta^4\omega^2}{R\kappa}
\right\}\right] -\frac{3\kappa
U[\tilde{\mu}\omega+\tilde{q}(1+\omega^2)]C'}{2EC(1-\omega^2)(1-\delta^4R^{-2})}.
\end{align}
The second of above equation, after making some lengthy
calculations, provides
\begin{align}\nonumber
\mathcal{E}'+\frac{3C'}{C(1-\omega^2)}\mathcal{E}&=\frac{\kappa\tilde{\mu}'}{2(1-\delta^4R^{-2})}
-\frac{3C'}{C(1-\omega^2)(1-\delta^4R^{-2})}
\left[\frac{\kappa\omega\tilde{\mu}U}{2E}-\frac{\omega^2\delta^4}{R}\right.\\\label{67}
&+\left.\kappa
\tilde{q}\left\{\omega+\frac{U(1+\omega^2)}{2E}\right\} \right],
\end{align}
from which, we have obtained the constraint on heat conducting
scalar as follows
\begin{align}\label{68}
\tilde{q}&=\frac{\omega}{2\kappa}\left(\frac{\kappa
U\tilde{\mu}}{E}-\frac{\omega\delta^4}{R}\right)\left\{\omega+\frac{U(1+\omega^2)}{2E}\right\}^{-1}.
\end{align}
The dissipative dust with tilted congruences will be of regular
energy density, if $\mu'$, Weyl scalar as well as above value of
heat flux is zero. This clearly shows that homogeneity depends upon
dark source Palatini $f(R)$ terms and congruence radial velocity
$\omega$.

\section{Summary}

We have seen that LTB spacetimes as seen by a tilted observer
exhibit physical properties which drastically differ from those
present in the standard non-tilted LTB.

In this paper, we have studied the dynamics of LTB anisotropic
geometry from the point of view of a tilted observer in Palatini
$f(R)$ gravity. The non-ideal matter distribution and the congruence
supported by its 4-velocity vector, as observed by the tilted
observer, is analyzed in detail. The ``inhomogeneity factor", i.e.,
the variable quantities depicting those aspects of the matter
configurations that are involved in the emergence of energy-density
irregularities, has been explored with respect to the tilted
congruence. We have also integrated evolution of such factor in the
maintenance of homogeneous phases of compact objects.

The dynamical equations and kinematical quantities are explored in
non-comoving coordinates for the systematic construction of our
analysis. Two expressions widely known to be Ellis equations have
been developed in the context of Palatini $f(R)$ gravity. These
equations have linked the Weyl tensor with the material variables as
seen by the tilted observer. We have extracted the factors that are
responsible for the emergence of inhomogeneities in the LTB energy
density under particular cases of dissipative and non-dissipative
regimes. In non-radiating sector, we studied irregularity factors
for the cloud of non-interacting particles, isotropic fluid and
anisotropic matter while the radiating sector is explored only for
non-interacting particles in tilted frame. The results in these
particular cases are summarized as follows.
\begin{itemize}
\item For non-interacting and non-dissipative particles, we observed that
the initially homogeneous system will remain homogeneous if it is
conformally flat or have zero expansion scalar. It means that the
inhomogeneity in the LTB type universe is not only controlled by the
Weyl tensor but also the expansion scalar. We would like to stress
here that the converse is not true, i.e., zero expansion condition
does not lead to a homogeneous density distribution. It is worth
mentioning that expansion-free scenarios have their own physical
interpretation during the evolutionary process with some crucial
impact on realistic models that we mentioned earlier in subsection
4.1.1.
\item With the inclusion of isotropic pressure in the non-interacting
particles, we found that the effects on the inhomogeneity parameters
of density distribution differs from the previous one. We observed
that a geometrical combination of dark source terms of Palatini
$f(R)$ gravitational field $\Psi$ along with Weyl tensor and
expansion scalar are the responsible factors. In the absence of
extra curvature invariants of the theory, the Weyl tensor will be
the only candidate for the appearance of inhomogeneities in the
density distribution. Furthermore, we have explored that if during
evolution, the expansion scalar is able to attain a specific value
(Eq.(\ref{58})), then the system will have regular environment of
energy density.
\item Similar factors for the case of non-radiating anisotropic
matter distribution are obtained. For the smooth distribution of
energy density, the value of $\Theta$ has also been identified
(mentioned in Eq.(\ref{62})). The corresponding inhomogeneity factor
$\Phi$ has also been explored (Eq.(\ref{64})). It is seen that
Palatini $f(R)$ dark source terms and tilted parameter $\omega$ have
produced hindrances for the system to leave initial homogeneous
state of the compact object.
\item In the radiating dust cloud case, we observed that a specific
value of dissipation obtained in Eq.(\ref{68}) is the responsible
factor of density inhomogeneity in Palatini $f(R)$ gravity and
tilted observer along with the Weyl tensor. The homogeneous state
can be recovered if the system is conformally flat and
non-radiating.
\end{itemize}

All of our results support the analysis of \cite{zb11w} on setting
$\omega=0$, while the assumptions $\omega=0$ and $f(R)=R$, in our
calculations would provide results compatible with \cite{11aa} and
\cite{zb3}.

\vspace{0.3cm}

\section*{Acknowledgment}

This work was partially supported by the JSPS KAKENHI Grant Number
JP 25800136 and the research-funds presented by Fukushima University
(K.B.).

\vspace{0.3cm}

\renewcommand{\theequation}{A\arabic{equation}}
\setcounter{equation}{0}
\section*{Appendix}

The parts of Eqs.(\ref{39}) and (\ref{40}) are
\begin{align}\nonumber
\mathcal{D}_0&=-\dot{\mathcal{T}}_{00}+\left(\frac{\mathcal{T}_{10}}{B^2}\right)'
+\frac{\mathcal{T}_{01}}{B^2}\left(\frac{2f'_R}{f_R}+\frac{B'}{B}+\frac{2C'}{C}\right)
-\mathcal{T}_{00}\left(\frac{\dot{B}}{B}+\frac{3\dot{f_R}}{2f_R}
+\frac{2\dot{C}}{C}\right)\\\nonumber
&-\frac{2\mathcal{T}_{22}}{C^2}\left(\frac{\dot{C}}{C}
+\frac{\dot{f_R}}{2f_R}\right)-\frac{\mathcal{T}_{11}}{B^2}\left(\frac{\dot{B}}{B}+\frac{\dot{f_R}}
{2f_R}\right),\\\nonumber
\mathcal{D}_1&=\mathcal{T}_{00}\frac{f_R'}{2f_R}-\dot{\mathcal{T}}_{10}+\left(\frac{\mathcal{T}_{11}}{B^2}\right)'
-\mathcal{T}_{10}
\left(\frac{2\dot{f}_R}{f_R}+\frac{\dot{B}}{B}+\frac{2\dot{C}}{C}\right)
-\frac{2\mathcal{T}_{22}}{C^2}\left(\frac{C'}{C}+\frac{f'_R}{2f_R}\right)\\\nonumber
&+\frac{\mathcal{T}_{11}}{B^2}
\left(2\frac{C'}{C}+\frac{3f_R'}{2f_R}\right).
\end{align}


\vspace{0.5cm}

\begin{thebibliography}{40}

\bibitem{1} S. Perlmutter, et  al. (Supernova Cosmology Project Collaboration),
Astrophys. J. \textbf{517}, 565 (1999).

\bibitem{R-DE-MG}
%
S.~Nojiri and S.~D.~Odintsov,
Phys.\ Rept.\ {\bf 505}, 59 (2011)
[arXiv:1011.0544 [gr-qc]];
%
S.~Nojiri and S.~D.~Odintsov,
eConf C {\bf 0602061} (2006) 06
[Int.\ J.\ Geom.\ Meth.\ Mod.\ Phys.\ {\bf 4}, 115 (2007)]
[hep-th/0601213];
%
S.~Capozziello and V.~Faraoni,
\textit{Beyond Einstein Gravity}
(Springer, Dordrecht, 2010);
%
S.~Capozziello and M.~De Laurentis,
Phys.\ Rept.\ {\bf 509}, 167 (2011)
[arXiv:1108.6266 [gr-qc]];
%
  A.~de la Cruz-Dombriz and D.~S\'{a}ez-G\'{o}mez,
  Entropy {\bf 14}, 1717 (2012)
  [arXiv:1207.2663 [gr-qc]];
%
  K.~Bamba, S.~Capozziello, S.~Nojiri and S.~D.~Odintsov,
  Astrophys.\ Space Sci.\  {\bf 342}, 155 (2012)
  [arXiv:1205.3421 [gr-qc]];
%
  A.~Joyce, B.~Jain, J.~Khoury and M.~Trodden,
  Phys.\ Rept.\  {\bf 568}, 1 (2015)
  [arXiv:1407.0059 [astro-ph.CO]];
%
  K.~Koyama,
  Rept.\ Prog.\ Phys.\  {\bf 79}, 046902 (2016)
  [arXiv:1504.04623 [astro-ph.CO]];
%
  K.~Bamba and S.~D.~Odintsov,
  Symmetry {\bf 7}, 1, 220 (2015)
  [arXiv:1503.00442 [hep-th]].

\bibitem{2} L. Amendola, R. Gannouji, D. Polarski and S. Tsujikawa,
Phys. Rev. D \textbf{75}, 083504 (2007); G. Cognola, E. Elizalde, S.
Nojiri, S. D. Odintsov, L. Sebastiani and S. Zerbini, Phys. Rev. D
\textbf{77}, 046009 (2008); A. Starobinsky, J. Exp. Theor. Phys.
Lett. \textbf{86}, 157 (2007).

\bibitem{zb2a} G. J. Olmo, Phys. Rev. D \textbf{72}, 083505 (2005); Int. J. Mod.
Phys. D \textbf{20}, 413 (2011); Phys. Rev. D \textbf{86}, 044014
(2012); G. J. Olmo, H. Sanchis-Alepuz and S. Tripathi, Phys. Rev. D
\textbf{86}, 104039 (2012); S. Capozziello, T. Harko, T. S.
Koivisto, F. S. N. Lobo, and G. J. Olmo, Phys. Rev. D \textbf{86},
127504 (2012).

\bibitem{3} B. Li and M. -C. Chu, Phys. Rev. D
\textbf{74}, 104010 (2006).

\bibitem{4} K. Kainulainen,
J. Piilonen, V. Reijonen and D. Sunhede, Phys. Rev. D \textbf{76},
024020 (2007).

\bibitem{5} S. Fay, R. Tavakol and S. Tsujikawa, Phys. Rev. D \textbf{75}, 063509 (2007).

\bibitem{6} F. Shojai and A. Shojai, Phys. Rev. D \textbf{78}, 104011 (2008).

\bibitem{7} T. P. Sotiriou and V. Faraoni, Rev. Mod. Phys. \textbf{82}, 451 (2010).

\bibitem{8} Y. Kucukakca and U. Camci, Astrophys. Space Sci. \textbf{338}, 338 (2012).

\bibitem{8a} R. Penrose and S. W. Hawking, General Relativity, An Einstein
Centenary Survey (Cambridge University Press, 1979).

\bibitem{8aa} P. S. Joshi and
I. H. Dwivedi, Phys. Rev. D \textbf{47}, 5357 (1993).

\bibitem{9} D. M. Eardley and L. Smarr, Phys. Rev. D
\textbf{19}, 2239 (1979); R. Bali and A. Tyagi, Astrophys. Space
Sci. \textbf{173}, 233 (1990).

\bibitem{9a} L. Herrera, A. Di Prisco, J. L. Hernandez-Pastora and N. O. Santos,
Phys. Lett. A \textbf{237}, 113 (1998).

\bibitem{zb99} L. Herrera, A. Di Prisco, J. Ospino and J. Carot, Phys. Rev. D \textbf{91}, 024010
(2015); ibid., Phys. Rev. D \textbf{94}, 064072 (2016); J. L.
Hernandez-Pastora, L. Herrera and J. Martin, Class. Quantum Grav.
\textbf{33}, 235005 (2016) [arXiv:1607.02315 [gr-qc]].

\bibitem{10} F. C. Mena, B.C. Nolan, and R. Tavakol, Phys. Rev. D \textbf{70}, 084030 (2004).

\bibitem{10a} A. Di Prisco, L. Herrera, G. L. Denmat, M. A. H. MacCallum and
N. O. Santos, Phys. Rev. D \textbf{76}, 064017 (2007).

\bibitem{11} C. H. Chuang, J. N. Gu and W. Y. P. Hwang, Class. Quantum
Grav. \textbf{25}, 175001 (2008).

\bibitem{11a} L. Herrera, A. Di Prisco, E. Fuenmayor and O. Troconis, Int. J.
Mod. Phys. D \textbf{18}, 129 (2009).

\bibitem{11aa} L. Herrera, Int. J. Mod. Phys. D \textbf{20}, 1689 (2011) [arXiv:1101.1514 [gr-qc]].

\bibitem{zb11a} M. Z. Bhatti and Z. Yousaf, Eur. Phys. J. C
\textbf{76}, 219 (2016) [arXiv:1604.01395 [gr-qc]]; Z. Yousaf and M.
Z. Bhatti, Mon. Not. R. Astron. Soc. \textbf{458}, 1785 (2016); Z.
Yousaf and M. Z. Bhatti,  Eur. Phys. J. C  \textbf{76}, 267 (2016)
[arXiv:1604.06271 [physics.gen-ph]]; Z. Yousaf, M. Z. Bhatti and U.
Farwa, Mon. Not. R. Astron. Soc. \textbf{464}, 4509 (2016); M. Z.
Bhatti and Z. Yousaf, Int. J. Mod. Phys. D \textbf{26}, 1750029
(2017).

\bibitem{zb11q} Z. Yousaf, K. Bamba and M. Z. Bhatti, Phys. Rev. D
\textbf{93}, 064059 (2016) [arXiv1603.03175 [gr-qc]].

\bibitem{zb11w} Z. Yousaf, K. Bamba and M. Z. Bhatti, Phys. Rev. D
\textbf{93}, 124048 (2016) [arXiv1606.00147 [gr-qc]].

\bibitem{12} A. R. King and G. F. R. Ellis, Commun. Math. Phys. \textbf{31}(1973)209;
G. F. R. Ellis and A. R. King, Commun. Math. Phys. \textbf{38}, 119
(1974); C. B. Collins and G. F. R. Ellis, Phys. Rep. \textbf{56}, 65
(1979).

\bibitem{13} R. Bali and K. Sharma, Astrophys. Space Sci. \textbf{271}, 227 (2000);
R. Bali, and B. L. Meena, Astrophys. Space Sci. \textbf{281}, 565
(2002).

\bibitem{14} C. G. Hewitt, R. Bridson and J.
Wainwright, Gen. Relativ. Grav. \textbf{33}, 65 (2001).

\bibitem{15} D. D. Pawar, S. W. Bhaware and A. G. Deshmukh, Rom. J. Phys.
\textbf{54}, 187 (2009).

\bibitem{16} P. S. Apostolopoulos, Gen. Relativ. Gravit. \textbf{37}, 937 (2005).

\bibitem{17} S. K. Sahu and T. Kumar, Int. J. Theor. Phys. \textbf{52}, 793 (2013).

\bibitem{18} M. Sharif and M. Z. Bhatti, Mod. Phys. Lett. A
\textbf{29}, 1450165 (2014); Int. J. Mod. Phys. D \textbf{24},
1550014 (2015); J. Exp. Theor. Phys. \textbf{120}, 813 (2015).

\bibitem{18inho} M. Sharif and Z. Yousaf, Eur. Phys. J. C \textbf{75}, 58
(2015); Astrophys. Space Sci. \textbf{357}, 49 (2015); Can. J. Phys.
\textbf{93}, 905 (2015); Gen. Relativ. Gravit. \textbf{47}, 48
(2015).

\bibitem{18co} M. Sharif and Z. Yousaf, Eur. Phys. J. C  \textbf{75}, 194 (2015) [arXiv:1504.04367v1
[gr-qc]]; Astrophys. Space Sci. \textbf{355}, 317 (2015); Int. J.
Theor. phys. \textbf{55}, 470 (2016).

\bibitem{zb2} K. Kainulainen, V. Reijonen and D. Sunhede, Phys. Rev. D \textbf{76}, 043503
(2007); G. J. Olmo, Phys. Rev. D \textbf{78}, 104026 (2008).

\bibitem{zb1} A. D. Felice and S. Tsujikawa, Living Rev. Relativity \textbf{13}, 3
(2010).

\bibitem{zb3} L. Herrera, A. Di Prisco and J. Iba\~{n}ez, Phys.
Rev. D \textbf{84}, 064036 (2011).

\bibitem{zb4} F. S. Labini, Class. Quantum Grav. \textbf{28},
164003 (2011).

\bibitem{zb10} S. M. Carroll, V. Duvvuri, M. Trodden and M. S. Turner, Phys. Rev. D
\textbf{70}, 043528 (2004).

\bibitem{zb11} C. W. Misner and D. Sharp, Phys. Rev. \textbf{136}, B571
(1964).

\bibitem{zb12} G. F. R. Ellis, Gen. Relativ. Gravit. \textbf{41}, 581 (2009).

\bibitem{zb13} L. Herrera, A. Di Prisco, J. Mart\'{i}n,
J. Ospino, N. O. Santos and O. Troconis, Phys. Rev. D \textbf{69},
084026 (2004).

\bibitem{zb14} P. S. Joshi, N. Dadhich, and R. Maartens, Phys. Rev. D \textbf{65},
101501 (2002).

\bibitem{zb15} K. S. Virbhadra, D. Narasimha, and S. M.
Chitre, Astron. Astrophys. \textbf{337}, 1 (1998).

\bibitem{zb16} K. S. Virbhadra and G. F. R. Ellis, Phys. Rev. D \textbf{65}, 103004
(2002).

\bibitem{zb6} M. Sharif and Z. Yousaf,
Astrophys. Space Sci. \textbf{352}, 943 (2014).

\bibitem{zb8} A. A. Starobinsky, Phys. Lett. B \textbf{91}, 99 (1980).


\end{thebibliography}
\end{document}